\documentclass[apj,twocolumn]{emulateapj}

\renewcommand\email\texttt

\shorttitle{The discovery of two globular clusters}
\shortauthors{Koposov et al.}
\slugcomment{Accepted for publication in ApJ} 
\begin{document}

\title{The discovery of two extremely low luminosity Milky Way
  globular clusters}

\author{S. Koposov\altaffilmark{1,2}, J.T.A. de Jong\altaffilmark{1},
  V. Belokurov\altaffilmark{2}, H.-W. Rix\altaffilmark{1},
  D.B. Zucker\altaffilmark{2}, N.W. Evans\altaffilmark{2},
  G. Gilmore\altaffilmark{2}, M.J. Irwin\altaffilmark{2},
  E.F. Bell\altaffilmark{1}}

\altaffiltext{1}{Max Planck Institute for Astronomy, K\"onigstuhl 17,
  69117, Heidelberg, Germany;\email{koposov,dejong@mpia.de}}
\altaffiltext{2}{Institute of Astronomy, University of Cambridge, Madingley Road, Cambridge CB3 0HA, UK}

\begin{abstract}
We report the discovery of two extremely low luminosity globular
clusters in the Milky Way Halo. These objects were detected in the Sloan
Digital Sky Survey Data Release 5 and confirmed with deeper imaging at the Calar
Alto Observatory. The clusters, Koposov 1 and Koposov 2, are located at
$\sim 40-50$ kpc and appear to have old stellar populations and luminosities of
only $M_V \sim -1 mag$. Their observed sizes of $\sim 3$ pc are well within the
expected tidal limit of $\sim$10 pc at that distance. Together with Palomar 1,
AM 4 and Whiting 1, these new clusters are the lowest luminosity globulars
orbiting the Milky Way, with Koposov 2 the most extreme. Koposov 1 appears to
lie close to distant branch of the Sagittarius stream. The half-mass relaxation
times of Koposov 1 and 2 are only $\sim 70$ and $\sim 55$ Myr respectively
(2 orders of magnitude shorter than the age of the stellar populations), so it
would seem that they have undergone drastic mass segregation. Since they do not
appear to be very concentrated, their evaporation timescales may be as low as
$\sim 0.1 t_{\rm Hubble}$. These discoveries show that the structural parameter
space of globular clusters in the Milky Way halo is not yet fully
explored. They also add, through their short remaining survival times,
significant direct evidence for a once much larger population of
globular clusters.
\end{abstract}


\section{Introduction}

The population of globular clusters around the Milky Way has been
studied extensively and the current census finds the majority at low
latitudes in the inner Galaxy ($R_{\rm GC} < 20$ kpc). Globular
clusters are almost universally ``old'' [$t_{\rm age} \approx 0.5 -- 1
\times t_{\rm Hubble}$], show no convincing evidence for dark matter,
and have characteristic luminosities of $10^5L_\odot$ $M_V \sim -8$)
and typical sizes of 3 pc. Yet, the observed range of structural
properties (e.g. mass, size, and concentration) is quite wide. This
range is of great interest, as it appears to be determined by a set of
astrophysical processes: the initial structure and orbit; subsequent
external processes, such as galactic tides and dynamical friction; and
ensuing mass segregation, evaporation, and core collapse~\citep[see,
  e.g.,][]{gnedin_ostriker,meylan97}.  Indeed, there has long been a
sense that the observed population of Galactic globular clusters
mainly reflects the subset of objects that could survive for $\sim
t_{\rm Hubble}$. In individual cases, there is clear evidence for
internal reshaping processes~\citep[as in \object{M15},][]{sosin97} and tidal
disruption~\citep[as in \object{Pal5},][]{odenkirchen}. Within this context,
identification and study of globular clusters with extreme properties is undoubtely of
great interest.

Our census of objects at the outskirts of the Milky Way has increased
rapidly in the last few years, mostly based on large-area CCD surveys
such as the Sloan Digital Sky Survey~\citep[SDSS;][]{york}. Recent
searches for Galactic halo objects have not only found many dwarf
galaxies~\citep{willman,canven,5pack,irwin}, but also added two faint
and extended objects that may be Milky Way globular clusters.  The newcomers,
\object{Willman 1} and \object{Segue 1}, both have distorted
irregular isopleths, perhaps indicating ongoing tidal disruption.
\object{Willman 1} seems to show some evidence for dark matter and metallicity
spread \citep{martin}, casting some doubt on whether it is a
globular cluster at all.

Here, we announce the discovery of two new, distant, extremely faint
and compact ($\sim 3$ pc) globular clusters, named \object{Koposov 1} and
\object{Koposov 2}, first detected in SDSS Data Release 5 (DR5) and
subsequently confirmed with deeper imaging at Calar Alto. The total
luminosity of Koposov 2 appears to be $\sim\ -1$ mag, lower than that of
the faintest Galactic globular known to date, \object{AM 4}
\citep[$-1.4$ mag,][]{am4}.  Koposov 1 is not much brighter; at
$M_{V,tot} \sim -2$ mag, it has the third-lowest luminosity. In total,
only 3 out of the previously known $\sim 160$ Galactic clusters, have
comparably low luminosity and small sizes: AM 4, \object{Palomar 1}, and
\object{Whiting 1} \citep{whiting}. Willman 1 and Segue 1 also have extremely
low luminosities, but are an order of magnitude larger.

In this paper, we describe the deep follow-up data confirming the
discoveries and give estimates of the structural parameters of the new objects (see Table~\ref{prop_table}).
We argue that the discovery of these two low-mass globulars in less than 1/5 of
the sky may mean that a substantial population of such clusters lurks in
the outer halo of the Milky Way.

\section{Discovery and observations}

\begin{figure*}
\plottwo{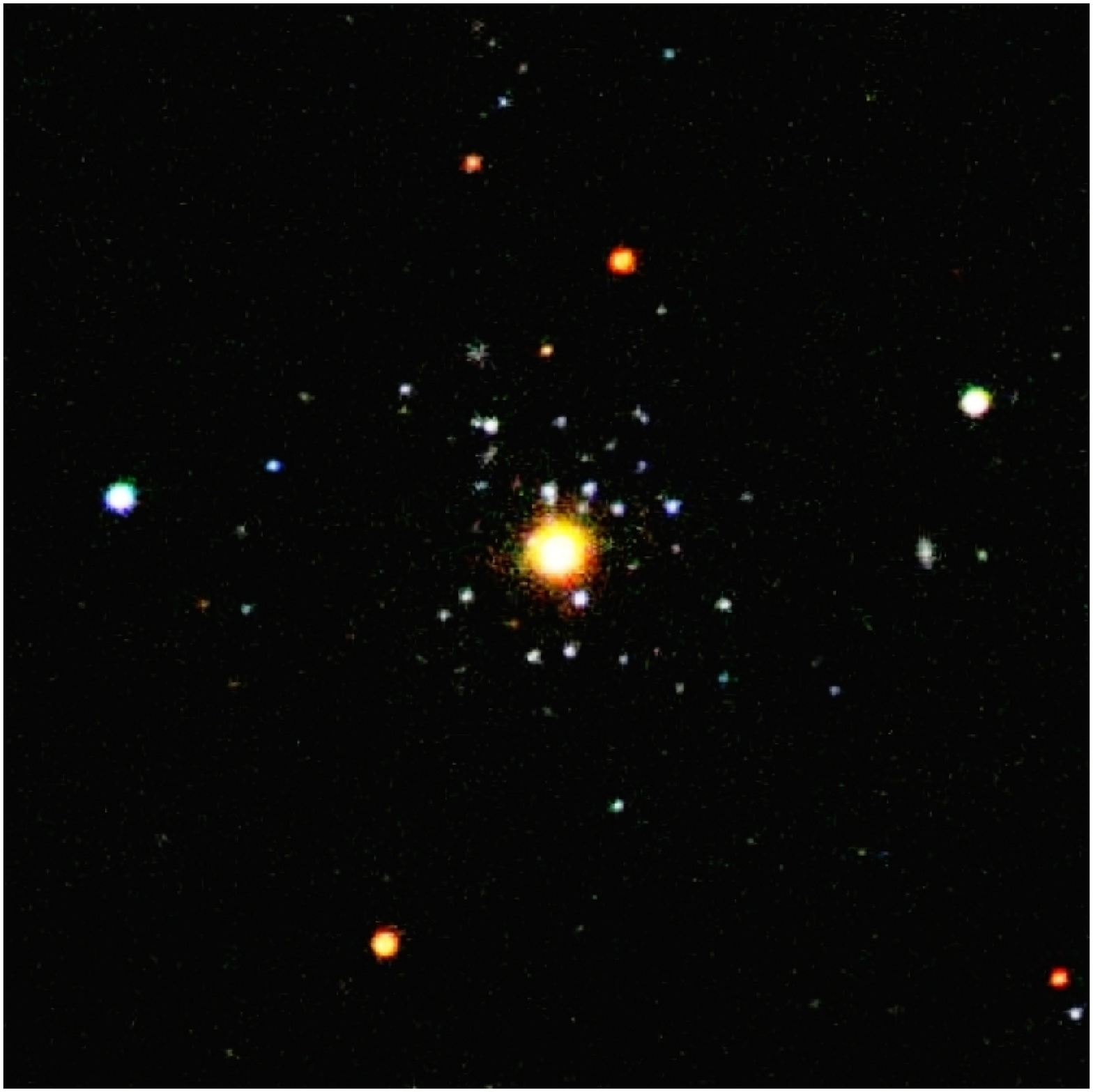}{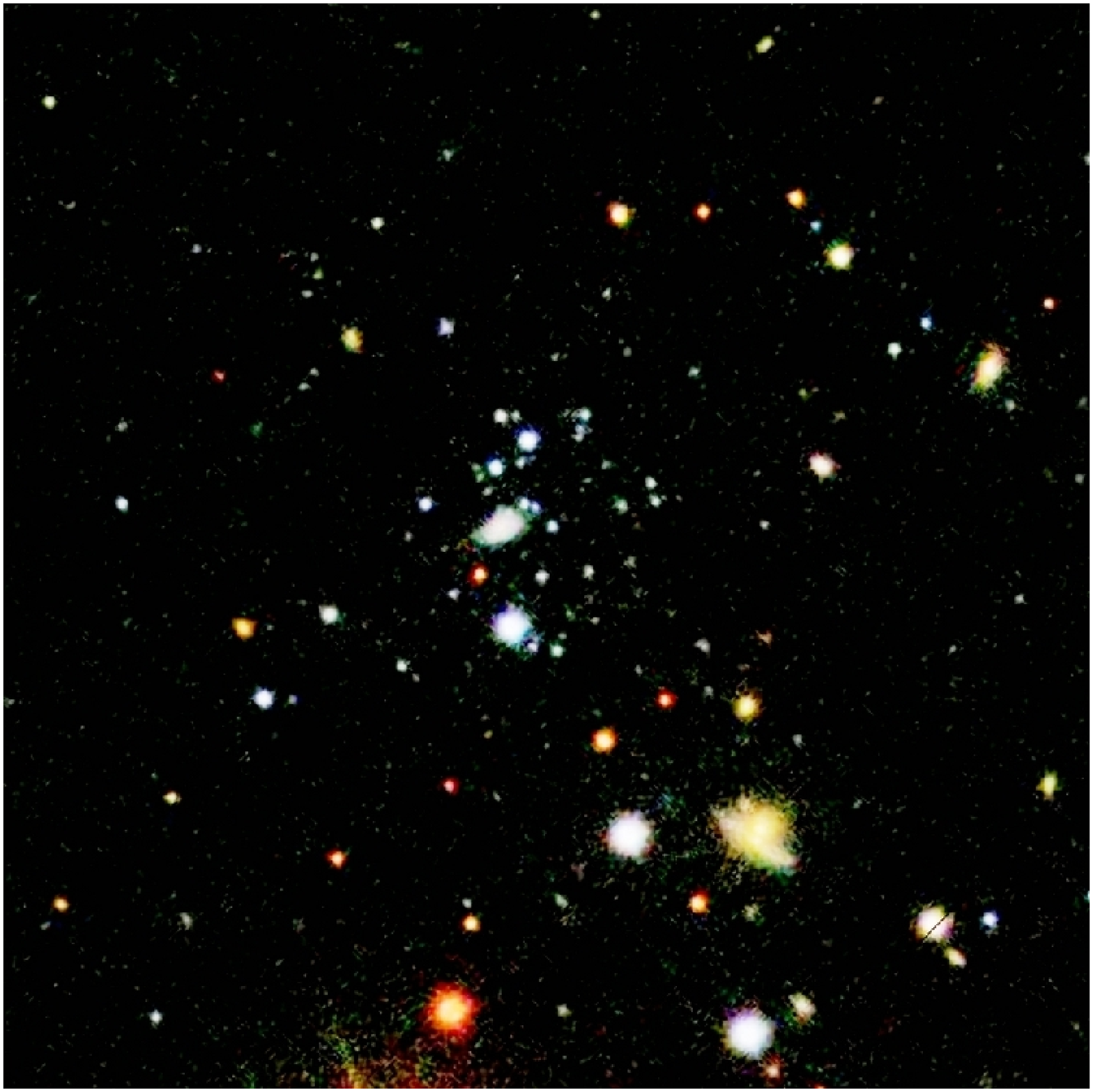}
\caption{$3\arcmin \times 3\arcmin$ SDSS cutout images
of Koposov 1 and 2. The bright star in the
   center of Koposov 1 is a foreground star with $V\sim14.5^m$ and large
   proper motion $(\mu_\alpha,\mu_\delta )\sim(-32,-12 ) \mu as\,yr^{-1}$,
   according to the USNO-B1 catalog~\citep{usno}. The bright extended
   object near the center of Koposov 2 is a background galaxy.}
\label{sdss_cutouts}
\end{figure*}

\begin{deluxetable}{lcc}
 \tablecaption{Properties of Koposov 1 and Koposov 2}
 \tablecolumns{2}
\tablehead{\colhead{Parameter}&\colhead{Koposov 1}&\colhead{Koposov 2}}
 \startdata
 Coordinates (J2000) & 11:59:18.5 +12:15:36 & 07:58:17.00 +26:15:18 \\
 Coordinates ($\ell,b$) &  $(260.98\degr, 70.75\degr)$
 	& $(195.11\degr, 25.55\degr)$\\
Distance & $\sim 50$ kpc & $\sim 40$ kpc\\
Size & $\sim 3$ pc & $\sim 3$ pc\\
$M_V$  & $\sim -2^m$ & $\sim -1^m$ \\
Relaxation Time & $\sim 70$ Myrs & $\sim 55$ Myrs \\
Tidal radius & $\sim 11$ pc & $\sim 9$ pc\\ 
\enddata
\label{prop_table}
\end{deluxetable}

The two new globular clusters were originally selected among other
candidates in the course of our systematic search for small-scale substructure
in
the Milky Way halo. The aim of the search was to detect all
significant small-scale stellar overdensities above the slowly varying
Galactic background that are likely to be either dwarf spheroidal
galaxies or globular clusters. A detailed description of the algorithm
and its efficiency will be provided in a future paper, and we only
present here a brief outline of the method. The algorithm is based on
the so-called Difference of Gaussians method, first developed in
Computer Vision~\citep{babaud,lindenberg}. Starting from a
flux-limited catalog of stellar positions, the number-counts map in
$(\alpha,\delta)$ plane is convolved with a filter optimized for the
detection of overdensities, namely the difference of two two-dimensional Gaussians
\citep{koposov}. Having zero integral, the kernel guarantees that the
convolution with a constant (or slowly varying) background will result
in zero signal. When the data contain an overdensity with a size
comparable to the size of the inner Gaussian, the filter will be close
to optimal. 

\begin{figure*}
\plotone{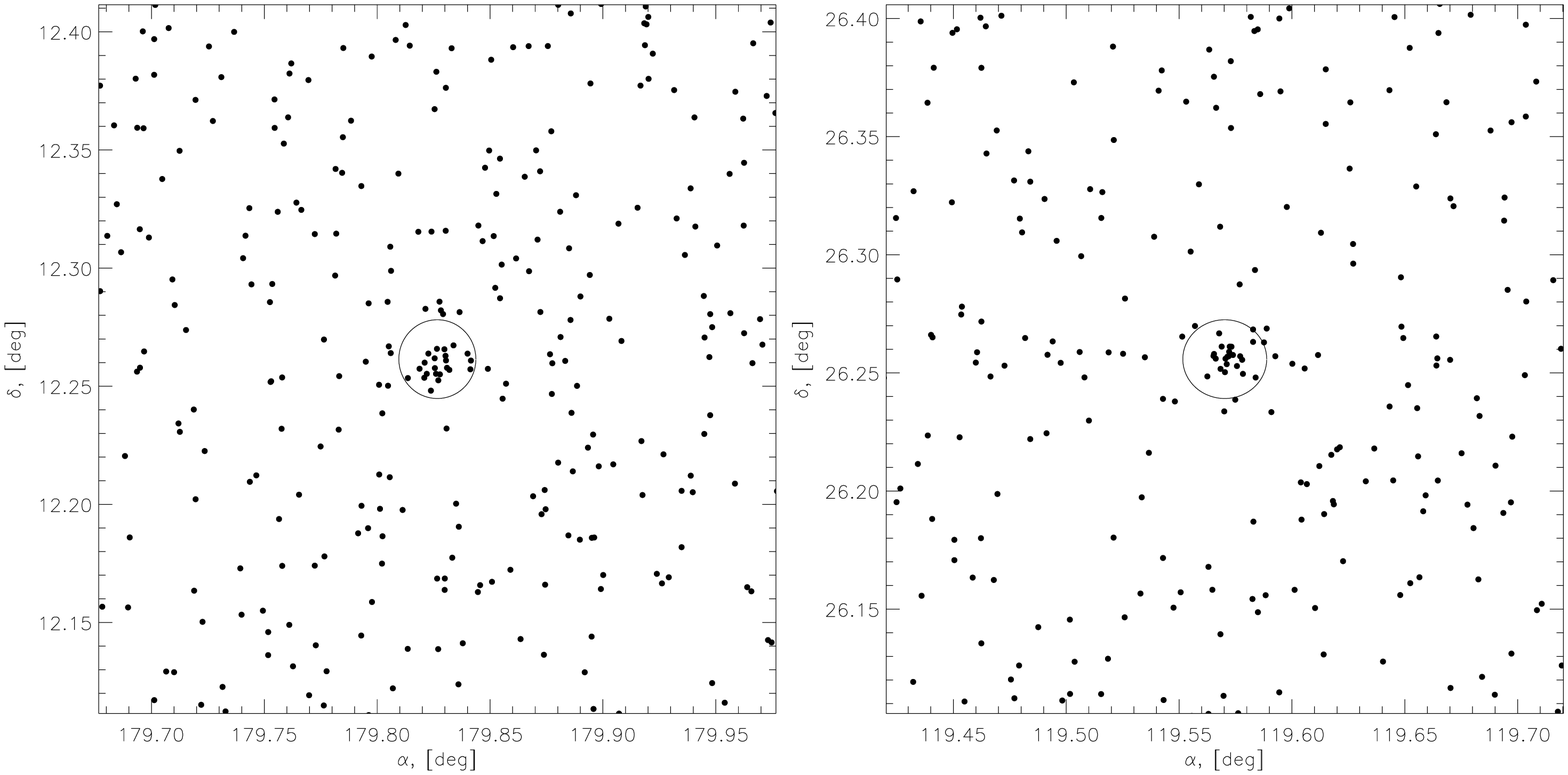}
\caption{The spatial distribution of the objects in the area of Koposov 1
and Koposov 2. All objects classified as stars with colors (g-r)$<0.6^m$ and
r$>20^m$ in the area $0.3\degr\times0.3\degr $ are shown. The circles with
1\arcmin\ radii centered on the objects are overplotted. 
   }
\label{sdss_dots}
\end{figure*}

We applied this filtering procedure to the entire stellar subset of
the DR5 source catalog with $r < 22^m, g-r < 1.2^m$. In our analysis we used
the photometry cleaned by switching on quality flags as described in SDSS SQL
pages~\footnote{\url{http://cas.sdss.org/dr5/en/help/docs/realquery.asp\#flags}}
This minimizes the influence of various artefacts including those caused by
proximity
of very bright or extended objects. In the resulting map that had been convolved
with a 2\arcmin\ kernel, we found two very compact objects among other
overdensities ranked highly according to their statistical significance.
Figure~\ref{sdss_cutouts} shows the SDSS images, and Figure~\ref{sdss_dots}
shows the spatial distribution of extracted sources, where central
concentrations of stars are clearly visible. 
These concentrations are detected at high level of significance. The areas of 
1\arcmin\ radius marked by circles centered on Koposov 1 and 2 plotted in
 Figure~\ref{sdss_dots} contain 22 objects and 23 objects, respectively, while
mean density of g-r$<0.6$, r$>20 mag$ stars should produce approximately 2.5
objects, which
implies a high statistical significance of the overdensities; for pure
Poisson distribution of objects, the probability to find such group of stars in 
all DR5 is around $10^{-9}$.

\begin{figure*}
\plotone{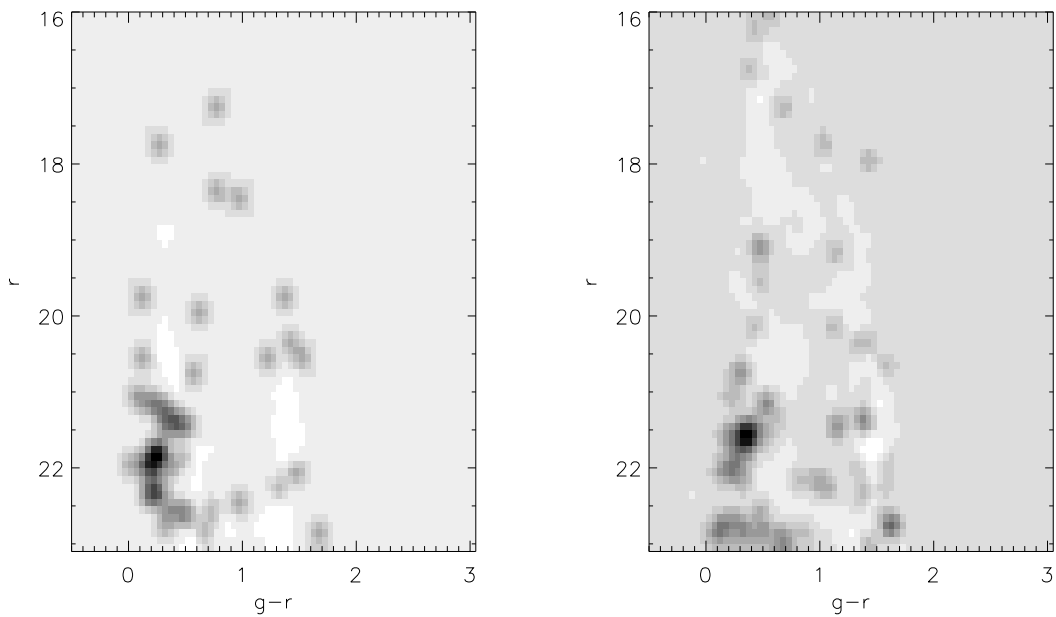}
\caption{The residual $g-r$ versus $g$ Hess diagrams of the clusters from
 the SDSS data. In each case the residual Hess diagram is costructed by
subtracting the normalized background Hess diagram  from the Hess diagram of
stars lying within 2\arcmin.5 radius from the centers of objects}
\label{sdss_hess}
\end{figure*}

The differential Hess diagrams for stars within $2\arcmin.5$ radius centered on
the objects are shown in Figure~\ref{sdss_hess}. There is a clear excess of blue
stars ($g-r < 0.5$), which we interpret as main-sequence turnoff stars at
$r\sim 22$, which roughly corresponds to distances of $\sim 50$kpc.

\begin{figure*} 
\plotone{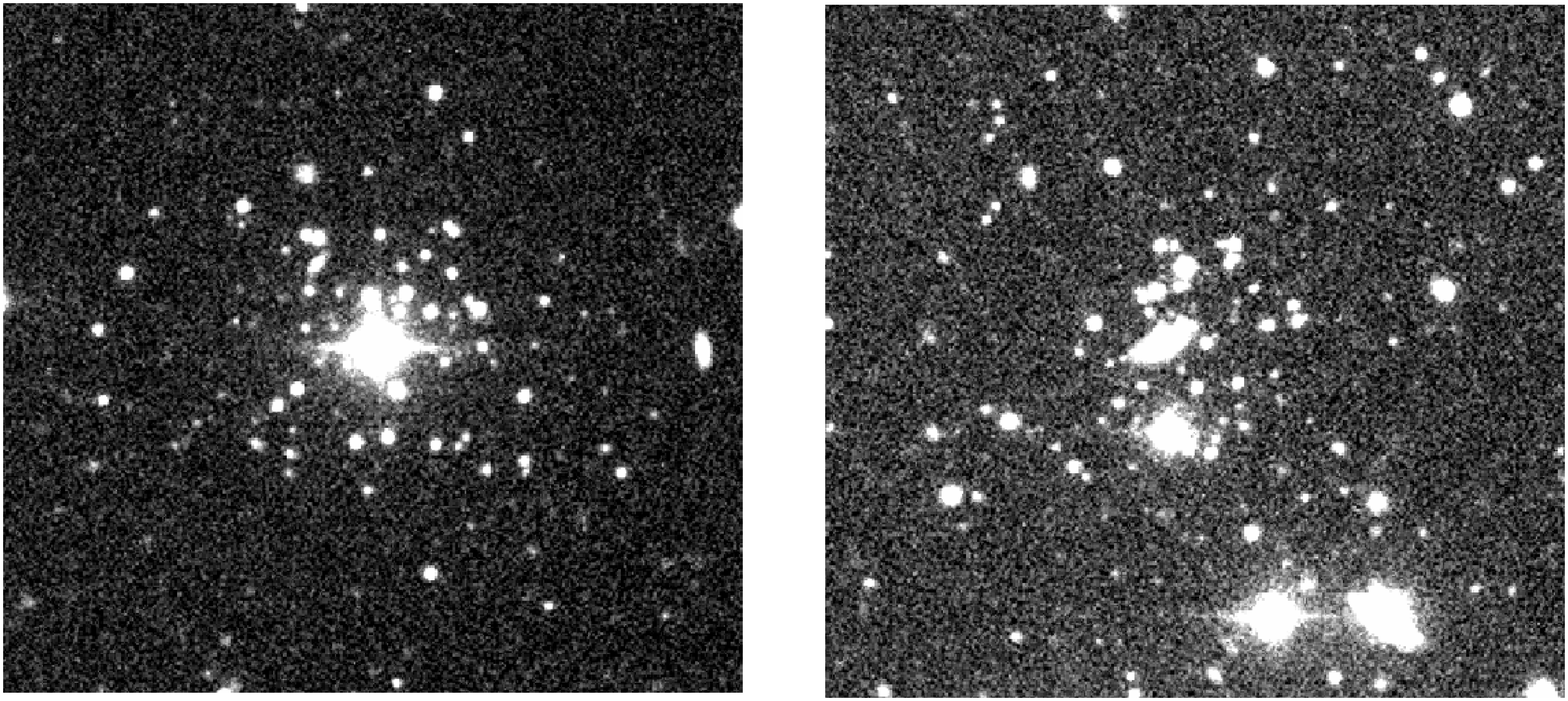}
\caption{$B$ band Calar Alto view of Koposov 1 and Koposov 2. The
   $2\arcmin \times 2\arcmin$ images are centered on the clusters
   (north is up, east is left).}
\label{ca_images}
\end{figure*}

To confirm the nature of discovered candidates and quantify their structural
and population properties, we acquired follow-up GTO observations in 2007 January
on the 2.2m telescope at Calar Alto using the CAFOS camera. This camera has
a $2k\times2k$ CCD with a $16\arcmin\times16\arcmin$ field of view and a pixel
scale of $0\arcsec.5\,pixel{^-1}$.  We observed each object for a total of 2 hr in
Johnson $B$ and 1.5 hr in Cousins $R$. The integrations were split into five
individual dithered exposures for cosmic ray and bad pixel
rejection. The observations were carried out in good photometric
conditions with a seeing of $1\arcsec-1.3\arcsec$. The data were
bias-subtracted and flat-fielded. The individual frames were
WCS-aligned, drizzled, and median-combined using our software and the
SCAMP and SWARP programs~\citep{scamp}. The combined $B$ band images of
the objects are shown in Figure~\ref{ca_images}.

\begin{figure*}
\plottwo{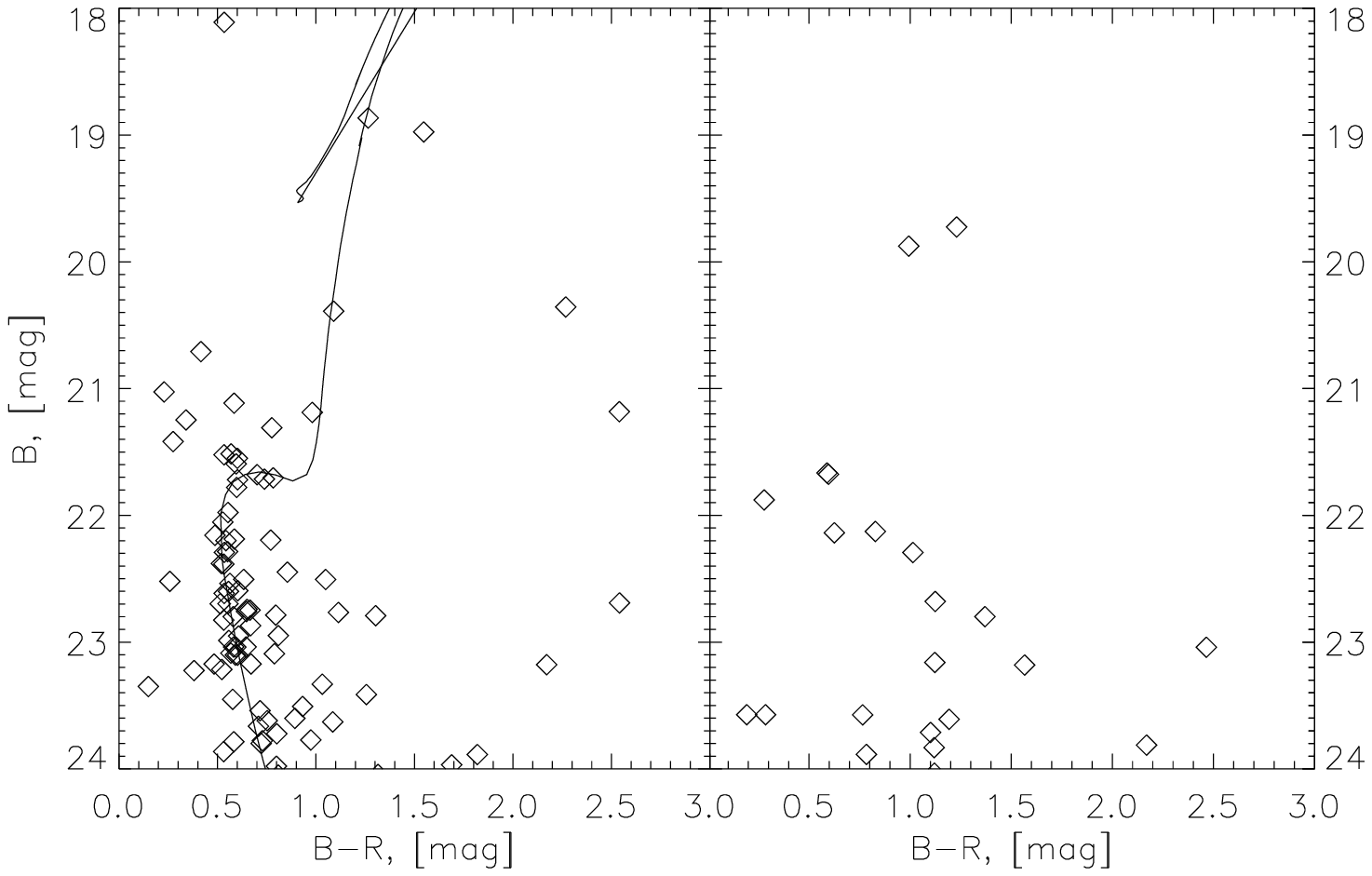}{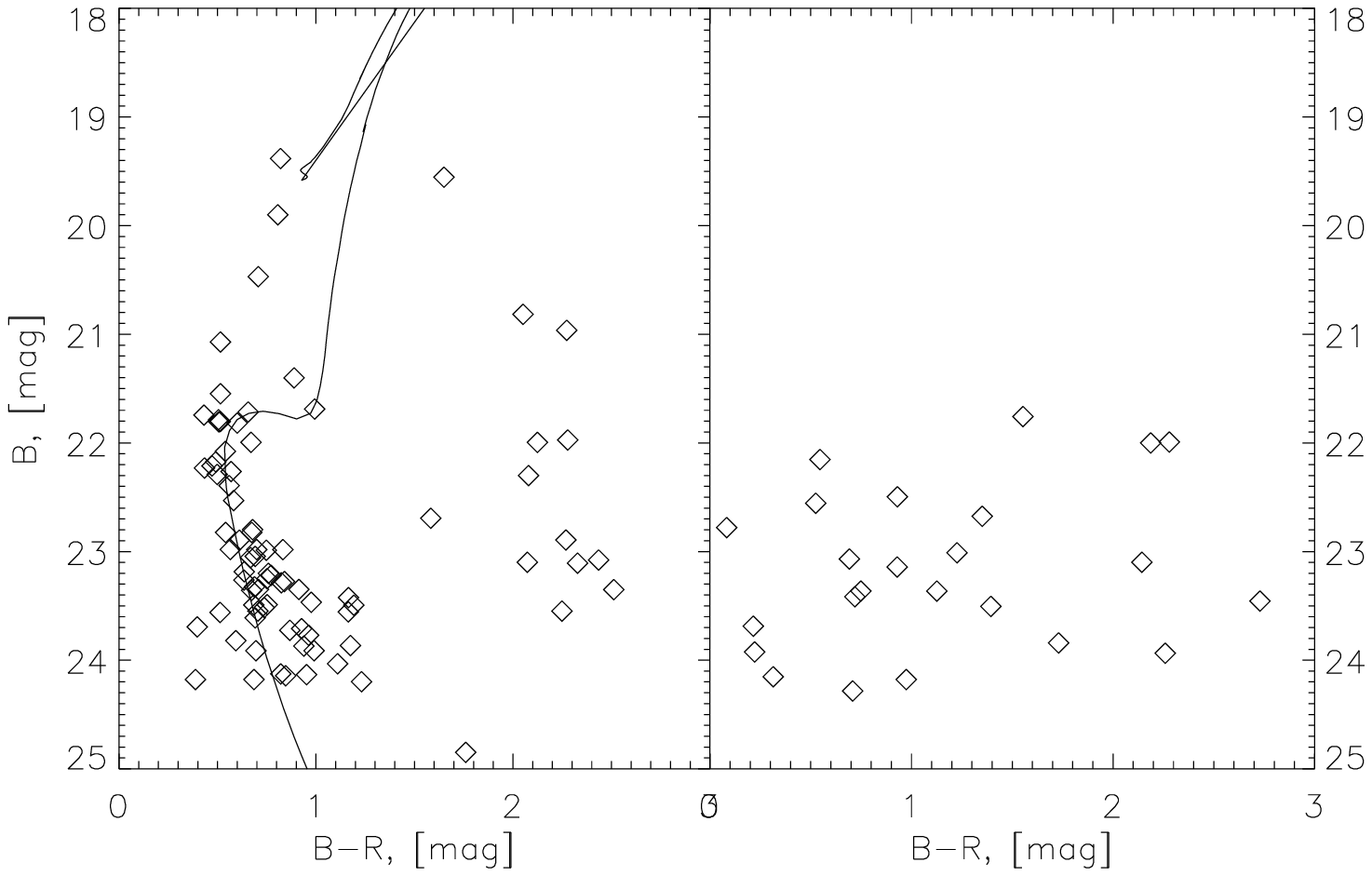}
\caption{\it Left panel, left half \rm: $B$ vs. $B-R$ CMDs derived from the
Calar Alto data for stars lying within 2\arcmin\ of Koposov 1 with 8 Gyr and
[Fe/H]=-2~\citet{girardi} isochrones overplotted. \it Left panel, right half\rm
: For comparison, the CMDs of stars in the annulus centered on Koposov 1 defined
by radii 3.2\arcmin\ and 3.7\arcmin are plotted. \it Right panel, left half\rm : $B$ vs.
$B-R$ CMDs of stars lying within
1.\arcmin2 of Koposov 2 with 8 Gyr and [Fe/H]=-2~\citet{girardi} isochrones
overplotted.\it Right panel, right half\rm : for comparison, the CMDs of stars
in the annulus centered on
Koposov 2 defined by radii 2\arcmin\ and 2.3\arcmin are plotted.}
\label{cmds}
\end{figure*}

The central stellar overdensities are clearly corroborated by the
Calar Alto photometry, which is nearly 2 mag deeper than the
original SDSS data. While the follow-up data are quite deep, the stars
are subject to significant crowding, due to the compactness of the
clusters. Therefore, for the purposes of robust source detection and
photometry, we used the DAOPHOT/ALLSTAR software \citep{daophot}. To
get the absolute calibration of the photometry from each frame, we
cross-matched the DAOPHOT sources with the SDSS catalog using the
Virtual Observatory resource SAI
CAS~\footnote{\url{http://vo.astronet.ru/cas} } \citep{sai_cas}. To
convert the Sloan $g$ and $r$ magnitudes into the Johnson-Cousins
photometric system, we used the conversion coefficients
from~\citet{smith}. The resulting $B$ versus
$B-R$ color-magnitude diagrams (CMDs) of the central regions of the objects together
with the CMDs of the comparison fields, extending to $B\sim
23.5^m-24^m$, are shown in the Figure~\ref{cmds}. The median photometric
accuracy of the data is  0.05-0.1 mag. The CMDs clearly show the presence of the main sequences near the centers of the
objects, while they are absent in the the comparison fields. The
statistical significance of the overdensities is also clearly supported by the new
data. The CMD of objects within 2\arcmin\ from the center of Koposov 1
contain 96 objects, while the background density inferred from the comparison
field should give around 23 objects, which gives a 15 $\sigma$ deviation. For
Koposov~2 , the number of objects within 1.2\arcmin\ is 92, while the background
density from the comparison field should produce around 24 objects, which gives
a 14~sigma deviation. In the next section we will discuss the properties of the
objects which can be derived from the follow-up data.

\section{Properties}
\label{properties_section}
The CMDs of the objects from the Calar
Alto data (Fig.~\ref{cmds}) clearly show a distribution of stars which can be
attributed quite convincingly to an old main sequence. In the case of Koposov 1,
the main-sequence turnoff is clear-cut, while for the second cluster it is not
so well defined. To estimate the distances to the objects, we
overplot in Figure~\ref{cmds} the 8 Gyr [Fe/H]=-2 isochrones from
\citet{girardi}. For Koposov 1, this gives a distance of 50 kpc. For Koposov 2,
the estimate is 40$\pm$5 kpc, but it is not well constrained due to a lack
of main-sequence turnoff stars. The angular diameters of the clusters are
$<0.5\arcmin$, which translates into a physical size of $r \sim
5$pc. Unfortunately, the number of stars detected in the central
regions is not enough to precisely measure half-light radii of the
objects; our best estimate is $r_h \sim 3$pc. For Koposov 1, we
subtracted the bright foreground star near the center, integrated the
light of the whole cluster in apertures and fitted it to a Plummer
profile with $r_h=3$pc. For Koposov 2, we performed a maximum
likelihood fit with $r_h \sim 3$pc.  Moreover, the minuscule number of
stars in both clusters does not allow us to  firmly establish their
total luminosities. Our estimate of $-1 \ga M_V \ga -2$ is based on the
absence of the giants in these clusters and the visible similarity of
the CMDs to that of the lowest luminosity globular cluster AM4
\citep[$M_V=-1.6$,][]{am4}. We checked that estimate by a simple Monte Carlo
experiment: using the Salpeter IMF and Girardi isochrones we simulated fake
clusters and deduced that the clusters with $-1\ga M_V \ga -2 $ have a number
of stars within 1.5-2 mag below the turnoff is close to the
observed number of stars (50-70) in our objects. We must say also that due to
the intrinsic faintness of the clusters and low number of stars in them the
estimates of the total luminosity and especially the age have large
uncertainties. However, with the existing data we cannot do any better. Much deeper
and more accurate photometry may be required to get precise age/luminosity
measures. The spectroscopic observations would be interesting in constraining
the metallicity of these objects, which is currently completely unknown.

We note that the CMD of Koposov 1 shows
several stars brighter and bluer than the tentative main-sequence turnoff, which
we interpret as blue stragglers. This hypothesis is not implausible
considering the low luminosity of the cluster and taking into account
the observed anti-correlation between the frequency of blue stragglers
and the luminosity of the globular cluster \citep{piotto}.

\begin{figure} 
\plotone{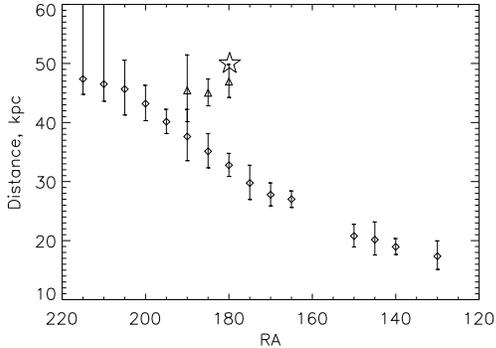}
\caption{Right ascension vs. distance for the A and C branches of
   the Sagittarius stream~\citep[see][]{field_of_streams}. The
   position of Koposov 1 is marked by a star.}
\label{sagi}
\end{figure}

The distance and the position of Koposov 1 suggest that this cluster
may be related to the Sagittarius tidal stream. Its location is a good
match to the distant tidal arm discovered in
\citet{field_of_streams}. Figure~\ref{sagi} shows the arms of the
Sagittarius stream in the DR5 slice around $\delta \sim 10\degr$ and
the position of Koposov 1.

\section{Discussion}

\begin{figure} 
\plotone{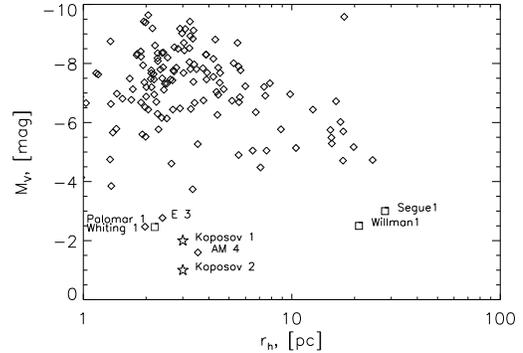}
 \caption{Size vs. absolute magnitude plot for Galactic globular
   clusters. The data from the \citet{harris} catalog are plotted with
   diamonds. Squares mark the locations of the recently discovered
   globular clusters Willman 1, Segue 1 and Whiting 1. Koposov 1 and 2
   are shown as stars.}
\label{size_mag}
\end{figure}

\begin{figure} 
\plotone{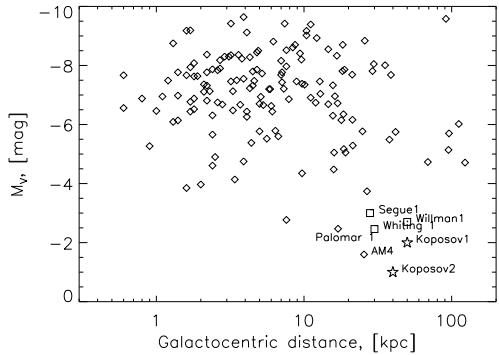}
\caption{Galactocentric distance vs. magnitude plot for
   Galactic globular clusters. Symbols are as in
   Fig.~\ref{size_mag}.}
\label{dist_mag}
\end{figure}

Figure~\ref{size_mag} shows Koposov 1 and 2 on the size-luminosity
plane along with other Galactic globular clusters. This illustrates
how unusual Koposov 1 and 2 are in their structural properties. It
appears that the detection of these clusters contributes to growing
evidence for a large population of small and extremely faint objects
(including Palomar 1, AM 4, E3 and Whiting 1). There is a
clear indication as well that this sub-population of globular clusters may
have significantly younger ages than classical globular clusters:
Palomar 1~\citep{sarajedini} and Whiting 1~\citep{whiting_new} have ages
between 4 and 6 Gyrs. The current estimate for the age of Koposov 1 is
$\approx 8$ Gyr, and the age of E3 globular cluster is $\approx$ 10 Gyr. This group
of clusters is also quite apparent on the galactocentric distance
versus luminosity plane shown in Figure~\ref{dist_mag}. At least 2
out of these 5 unusual clusters (Whiting 1 and Koposov 1) seem to be
associated with the Sagittarius dwarf galaxy.
Two quantities that are crucial for the long term evolution and
survival of Koposov 1 and 2 are the relaxation time and the expected
tidal radius.  For the half-mass relaxation time, we find
using equation (2-63) of \citet{spitzer} or equation (72) of \citet{meylan97},
$$t_{\rm rh} = 0.14 \frac{M_{\rm tot}^{1/2} {R_{\rm hl}}^{3/2}}{\langle m_*\rangle G^{1/2}
\ln(\Lambda)}=70\ {\rm and}\ 55\ {\rm Myr}$$
respectively for Koposov 1 and Koposov 2.
Here, we have assumed $L\approx 200 L_\odot$, $M/L \approx 1.5$,
$\langle m_* \rangle \approx 0.6 M_\odot$ and N=500 for Koposov 2,
while for Koposov 1, we have assumed twice as many stars, using the
observational estimates of \S~\ref{properties_section}. This means that
both clusters have extremely short relaxation times, less than 1\% of
$t_{\rm Hubble}$ and $t_{rh} \approx 0.01 t_{age,*}$. The most immediate effect
of two-body relaxation is mass segregation, which should be quite drastic given
the apparent stellar population age.  The
expected tidal truncation of these clusters occurs at~\citep[see,
e.g.,][]{innanen}
$$r_t = 0.43\left(\frac{M_{\rm cluster}}{M_{\rm MW}}\right)^{1/3}
\times R_{\rm peri} = 11\ {\rm and}\ 9\ {\rm pc}$$
where we have assumed an orbital eccentricity of 0.5, and that the
clusters are now near apocenter (hence $R_{peri} \approx 16$ kpc), a
Milky Way circular speed of 190 $km\,s^{-1}$ at 16 kpc and a cluster (stellar)
mass of 600 and 300 $M_{\odot}$ for Koposov 1 and 2, respectively. Hence, the
detectable extent of the globular clusters (3
pc) falls well within the tidal limit. From this argument, the
clusters are under no threat of destruction by tidal forces.  Although
formal profile fits are not feasible with so few stars, the stellar
distributions (see Figs.~\ref{sdss_cutouts} and \ref{ca_images}) are
well localized, but not centrally concentrated by globular cluster
standards; a core to tidal radius ratio of the observed stellar
distribution of 4 seems reasonable, implying a concentration parameter
of $c \equiv \log (r_t/r_c) \approx 0.5$. For such low concentrations,
the evaporation timescale $t_{\rm ev}$, which is the time-scale over
which two-body relaxation drives stars to beyond the escape velocity,
is $t_{\rm ev} \approx 1.5 t_{\rm cc} \approx 12 t_{\rm
rh}$(where $t_{\rm cc}$ is core collapse time)~\citep[Figure 17 and 19 in
][]{gnedin_lee_ostriker}For Koposov 1 and 2, this implies
evaporation time-scales of 0.7 and 1.1 Gyr, respectively. This estimate
of $t_{ev} \sim 0.1 t_{Hubble}$ may be an underestimate, if the brightest stars
which we observe are more concentrated than the faint stars due to mass
segregation; then the total mass and half-mass radius can be larger.
Nonetheless, this estimate makes it clear that the present structural and
dynamical state cannot have prevailed, even approximately, for a time-span of
$\sim$ 10 Gyr.
The above arguments hold irrespective of whether Koposov 1 and Koposov
2 were once part of a satellite galaxy, because they are mostly
derived from internal evolution factors. This discrepancy of
time-scales is more pronounced in Koposov 1 and 2, because their
relaxation time-scales are shorter than those of Palomar 1 and Whiting 1, which
in any case have accurate photometry suggesting younger ages of $\sim
4-6$\,Gyr.

At face value, Koposov 1 and 2 have survival times in their current
state of $\sim 0.1 t_{\rm Hubble}$, and were found in a search of 20\%
of the whole sky (SDSS DR5). The naive multiplication of these factors
points to a large parent population of $\sim 100$ objects.  The most
likely reservoir for this parent population is the globular clusters,
and possibly even old open clusters, in satellite galaxies that have
been accreted, like the Sagittarius. In objects like Koposov 1 and 2,
it is clear that the very short relaxation and evaporation times must
lead to drastic mass segregation and the expulsion of basically all
low-mass stars (this line of reasoning lead us to the modest $M/L
\approx 1.5$).  This gives new life to the view that truly
\textit{many} of the accreted globular clusters must have been
destroyed.  Yet, it is also clear that the actual dynamical prehistory
and future of these clusters requires much more careful modelling.
The small number of stars makes them ideal subjects of direct \textit{N}-body
calculations.  But regardless of their dynamical evolution, these
clusters manifestly demonstrate that the parameter space of globular
clusters in the Milky Way is not yet fully explored.

\acknowledgements{
  Based on observations collected at the Centro Astron\'omico Hispano
  AlemÃ\'an (CAHA) at Calar Alto, operated jointly by the Max-Planck
  Institut f\"ur Astronomie and the Instituto de Astrof\'i­sica de
  Andaluc\'ia (CSIC). Funding for the SDSS and SDSS-II has been provided
  by the Alfred P. Sloan Foundation, the Participating Institutions,
  the National Science Foundation, the U.S. Department of Energy, the
  National Aeronautics and Space Administration, the Japanese
  Monbukagakusho, the Max Planck Society, and the Higher Education
  Funding Council for England. The SDSS Web Site is
  http://www.sdss.org/. The SDSS is managed by the Astrophysical
  Research Consortium for the Participating Institutions. The
  Participating Institutions are the American Museum of Natural
  History, Astrophysical Institute Potsdam, University of Basel,
  Cambridge University, Case Western Reserve University, University of
  Chicago, Drexel University, Fermilab, the Institute for Advanced
  Study, the Japan Participation Group, Johns Hopkins University, the
  Joint Institute for Nuclear Astrophysics, the Kavli Institute for
  Particle Astrophysics and Cosmology, the Korean Scientist Group, the
  Chinese Academy of Sciences (LAMOST), Los Alamos National
  Laboratory, the Max-Planck-Institute for Astronomy (MPIA), the
  Max-Planck-Institute for Astrophysics (MPA), NewMexico State
  University, Ohio State University, University of Pittsburgh,
  University of Portsmouth, Princeton University, the United States
  Naval Observatory, and the University of Washington.  This research
  has made use of the SAI Catalog Access Services, Sternberg
  Astronomical Institute, Moscow, Russia.  S. Koposov is supported by
  the DFG through SFB 439 and by a EARA-EST Marie Curie Visiting
  fellowship.}



\begin{thebibliography}{}


\bibitem[Babaud et al.(1986)]{babaud} Babaud J., A. P. Witkin,
  M. Baudin, and R. O. Duda. 1986, IEEE Trans. Pattern Anal. Mach. Intell.,
8, 1, 26

\bibitem[Belokurov et al.(2006)]{field_of_streams} Belokurov, V., et 
al.\ 2006, \apjl, 642, L137 

\bibitem[Belokurov et al.(2007)]{5pack} Belokurov, V., et 
al.\ 2007, \apj, 654, 897 

\bibitem[Bertin(2006)]{scamp} Bertin, E.\ 2006, ASP 
Conf.~Ser.~351: Astronomical Data Analysis Software and Systems XV, 351, 
112

\bibitem[Carraro et al.(2007)]{whiting_new} Carraro G., A\&A,
  submitted, astro-ph/0702253

\bibitem[Girardi et al.(2000)]{girardi} Girardi, L., Bressan, 
A., Bertelli, G., \& Chiosi, C.\ 2000, \aaps, 141, 371 

\bibitem[Gnedin et al.(1999)]{gnedin_lee_ostriker} Gnedin, O.~Y., Lee, 
H.~M., \& Ostriker, J.~P.\ 1999, \apj, 522, 935 

\bibitem[Gnedin \& Ostriker(1997)]{gnedin_ostriker} Gnedin, O.~Y., \& 
Ostriker, J.~P.\ 1997, \apj, 474, 223 

\bibitem[Harris(1996)]{harris} Harris, W.~E.\ 1996, \aj, 112, 
1487 

\bibitem[Irwin et al.(2007)]{irwin} Irwin, M.~J., et al.\ 
2007, \apjl, 656, L13 

\bibitem[Inman \& Carney(1987)]{am4} Inman, R.~T., \& 
Carney, B.~W.\ 1987, \aj, 93, 1166 

\bibitem[Innanen et al.(1983)]{innanen} Innanen, K.~A., Harris, 
W.~E., \& Webbink, R.~F.\ 1983, \aj, 88, 338 


\bibitem[Koposov \& Bartunov(2006)]{sai_cas} Koposov, S.~E., \& 
Bartunov, O.~S.\ 2006, The Virtual Observatory in Action: New Science, New 
Technology, and Next Generation Facilities, 26th meeting of the IAU, 
SPS3, \#19, 3,  

\bibitem[Koposov et al.(2007)]{koposov} Koposov, S., et al.\ 
2007, ArXiv e-prints, 706, arXiv:0706.2687 


\bibitem[Lindenberg(1998)]{lindenberg} Lindenberg, T., 1998, International Journal of Computer Vision, 30, 2, 79

\bibitem[Martin et al.(2007)]{martin} Martin, N.~F., Ibata, 
R.~A., Chapman, S.~C., Irwin, M., \& Lewis, G.~F.\ 2007, \mnras, 380, 281 



\bibitem[Meylan \& Heggie(1997)]{meylan97} Meylan, G., \& 
Heggie, D.~C.\ 1997, \aapr, 8, 1

\bibitem[Monet et al.(2003)]{usno} Monet, D.~G., et al.\ 
2003, \aj, 125, 984 

\bibitem[Odenkirchen et al.(2001)]{odenkirchen} Odenkirchen, M., et 
al.\ 2001, \apjl, 548, L165


\bibitem[Piotto et al.(2004)]{piotto} Piotto, G., et al.\ 
2004, \apjl, 604, L109 

\bibitem[Sarajedini et al.(2006)]{sarajedini} Sarajedini, A., et 
al.\ 2006, ApJ, submitted, astro-ph/0612598 

\bibitem[Smith et al.(2002)]{smith} Smith, J.~A., et al.\ 
2002, \aj, 123, 2121 

\bibitem[Sosin \& King(1997)]{sosin97} Sosin, C., \& King, 
I.~R.\ 1997, \aj, 113, 1328 

\bibitem[Spitzer(1987)]{spitzer} Spitzer, L.\ 1987, Princeton, 
NJ, Princeton University Press, 1987

\bibitem[Stetson(1987)]{daophot} Stetson, P.~B.\ 1987, \pasp, 
99, 191 

\bibitem[Whiting et al.(2002)]{whiting} Whiting, A.~B., Hau, 
G.~K.~T., \& Irwin, M.\ 2002, \apjs, 141, 123 

\bibitem[Willman et al.(2005)]{willman} Willman, B., et al.\ 
2005, \apjl, 626, L85 

\bibitem[York et al.(2000)]{york} York, D.~G., et al.\ 2000, 
\aj, 120, 1579 

\bibitem[Zucker et al.(2006)]{canven} Zucker, D.~B., et al.\ 
2006, \apjl, 643, L103 


\end{thebibliography}
\end{document}